\documentclass[preprint,showpacs,floatfix]{revtex4}
\usepackage{natbib}
\usepackage{graphicx,subfigure,float,graphics}
\usepackage{epsfig}
\usepackage{subfig}
\usepackage{xcolor}
\usepackage{amssymb}
\usepackage{amsthm}
\usepackage{amsmath}
\usepackage{caption}
\usepackage{array,makecell}
\usepackage{multirow}

\usepackage[utf8]{inputenc}

\def\beq{\begin{equation}}
\def\eeq{\end{equation}}

\begin{document}

\title{Anomalous large-angle $\alpha$-scattering in a single-folding model with microscopic densities}

\author{A.-G. \c{S}erban$^{a}$, F. Salvat-Pujol$^{a}$, and N. S\u{a}ndulescu$^{b}$\footnote{corresponding author, email: sandulescu@theory.nipne.ro}}

\affiliation{ 
$^a$  European Organization for Nuclear Research, Esplanade des Particules 1, 1211 Geneva 23, Switzerland \\
$^b$  National Institute of Physics and Nuclear Engineering, 
077125 Bucharest-M\v{a}gurele, Romania}
\author{P. Marevi\'{c}}
\affiliation{
Department of Physics, Faculty of Science, University of Zagreb, Bijeni\v{c}ka c. 32, 10000 Zagreb, Croatia.}

\begin{abstract}

We investigate anomalous large-angle scattering (ALAS) of $\alpha$-particles from $N=Z$ nuclei within the framework of the single-folding model. Differential cross sections are calculated by folding the 
$\alpha$-nucleon interaction with nuclear density distributions obtained from both relativistic and non-relativistic mean-field models. The folding procedure employs a Gaussian-form $\alpha$-nucleon interaction, 
with its energy dependence and range constrained by previous theoretical studies. Our results show that ALAS in $sd$-shell nuclei is reasonably well reproduced using the microscopic densities together with an 
$\alpha$-nucleon interaction characterized by a unified parameter set, in which only two parameters
vary with the mass number.
   
\end{abstract}

\maketitle

\section{\label{sec:intro}Introduction}

The cross section of low-energy elastic $\alpha$-particle scattering from several light and medium-mass $N=Z$ nuclei exhibits a pronounced enhancement and an 
irregular energy dependence at large backward angles (see, \textit{e.g.}, Refs.~\cite{gaul,michel77,planeta} and references therein). These features, which differ 
from the diffractive behaviour observed at forward angles, are commonly referred to as anomalous large angle scattering (ALAS).

Since ALAS cannot be satisfactorily described within optical models based on a simple Woods-Saxon (WS) potential, a variety of interaction mechanisms have been 
proposed. These include $L$-dependent potentials~\cite{l-dep}, $\alpha$-exchange processes~\cite{exchange}, Regge-pole models~\cite{regge}, and quasi-molecular 
resonances~\cite{resonances}. Notably, ALAS is predominantly observed in scattering from $N = Z$ nuclei, which are expected to exhibit $\alpha$-like correlations in 
their ground states. This observation has motivated interpretations that associate ALAS with scattering between the incident $\alpha$ particle and preformed $\alpha$ 
clusters within the target nucleus~\cite{schmeing,michel_alpha,mahmoud}.

To describe ALAS, modified WS potentials have been also employed as an alternative to the standard optical model 
(see, \textit{e.g.}, Ref.~\cite{michel77}). While such 
phenomenological approaches can reproduce the experimental data, they rely on parameters that lack a clear microscopic interpretation. A more physically grounded 
framework is provided by the single-folding model (see, \textit{e.g.}, Refs.~\cite{gubler,kassem}), in which the interaction potential is obtained by folding the 
$\alpha$-nucleon interaction with the nuclear density distribution of the target.

Single-folding calculations have led to two main conclusions: (i) the real part of the scattering potential capable of describing ALAS deviates significantly from 
the standard WS form, and (ii) a satisfactory description of the data requires weak absorption of the incident $\alpha$ particles, typically implemented through an 
imaginary potential with small strength.

The connection between reduced absorption and ALAS has been discussed in Refs.~\cite{michel1,michel2,michel3}. According to these studies, ALAS arises from those 
$\alpha$ particles that, owing to weak absorption, penetrate into the target nucleus and are subsequently scattered back by the wall of the effective 
potential. This interpretation is supported by semiclassical calculations, which show that trajectories entering the nucleus are focused into a single focal point 
and that a fraction of them re-emerge at backward angles~\cite{michel3}.

In most folding calculations addressing ALAS, the real part of the optical potential is obtained by folding the $\alpha$-nucleon interaction with phenomenological 
density distributions of specific functional forms (\textit{e.g.}, modified Gaussian), adjusted to reproduce nuclear root-mean-square radii. Such densities are not expected 
to provide an accurate description of the density profile throughout the entire nucleus. In the present work, we instead employ realistic nuclear densities derived 
from self-consistent microscopic mean-field models. An advantage of this approach is that effects related to deformation and ground-state correlations of the target 
nuclei are naturally incorporated through the density distributions.

In previous studies of $\alpha$ scattering~\cite{gubler,kassem}, the single-folding model was used to determine only the real part of the optical potential, while 
the imaginary part was assumed to have a standard WS form. Here, both the real and imaginary parts of the optical potential are derived within the folding framework, 
using the same microscopic densities and $\alpha$-nucleon interaction. The latter is assumed to have a Gaussian form with explicit energy and density dependence. To 
limit the number of free parameters, the energy dependence and range of the real part are fixed according to theoretical estimates~\cite{lassaut}. We show that a 
reasonable description of ALAS in $sd$-shell nuclei can be achieved by allowing only the strengths of the real and imaginary components of the $\alpha$-nucleon 
interaction to vary with the mass number.

\section{Formalism}

\subsection{The folding model and the $\alpha$-nucleon interaction}

The differential cross section for elastic scattering of $\alpha$ particle on nuclei is calculated with the
following interaction potential between 
the $\alpha$ particle and the target nucleus:
\begin{equation}
\label{eq:total_pot}
    V(r_\alpha) = V_\mathrm{C}(r_\alpha) + V_{\alpha n} (r_\alpha),
\end{equation}
where $V_\mathrm{C}(r_\alpha)$ and $V_{\alpha n} (r_\alpha)$ are, respectively, the electrostatic and nuclear
interaction potentials, while $r_\alpha$ is the distance between the point-like $\alpha$ particle 
and the center of the target nucleus. 

The electrostatic potential is taken as the Coulomb interaction between a uniformly charged sphere 
(the target nucleus) and a point charge (the $\alpha$ particle). Its expression is:
\begin{equation}
\label{eq:coulomb_potential}
    V_\mathrm{C}(r_\alpha)= \begin{cases} \cfrac{Z_\mathrm{p} Z_\mathrm{t} e^2}{2 R}\left(3-\cfrac{r_\alpha^2}{R^2}\right),
                                          & r_\alpha < R, \\
                                          \cfrac{Z_\mathrm{p} Z_\mathrm{t} e^2}{r_\alpha},
                                          & r_\alpha \geq R,
                            \end{cases}
\end{equation}
where $R = 1.3 \hspace{0.025cm} A^{1/3}$ is the radius of the target nucleus with atomic mass $A$. $Z_\mathrm{p}$
and $Z_\mathrm{t}$ are the proton numbers of the projectile and of the target nucleus, respectively, and $e^2=1.44$~MeV~fm
is the square of the elementary charge.

The nuclear potential is obtained by folding an effective $\alpha$-nucleon interaction with the density distribution of the 
nucleons in the target nucleus (see, \textit{e.g.}, Ref.~\cite{satchler}). Specifically,
\begin{equation}
\label{eq:folding_potential}
    V_{\alpha n} (r_{\alpha}) = 4\pi \int v_{\alpha n} (s) \rho(r_n) r_n^2\mathrm{d}r_n,
\end{equation}
where $r_n$ denotes the coordinate of a nucleon inside the nucleus, $s= |r_{\alpha}-r_n|$ is the distance between the $\alpha$ particle and the nucleon, $\rho$ is the nuclear density distribution of the target nucleus, and 
$v_{\alpha n}$ represents the $\alpha$-nucleon interaction.

Within the single-folding model of Eq.~\eqref{eq:folding_potential}, the $\alpha$-nucleon interaction
is taken in the following phenomenological form:
\beq
  v_{\alpha n} (s) = V e^{-(s/a_\text{R})^2} + \mathrm{i} W  e^{-(s/a_\text{I})^2} ,
\eeq
where $V$ and $W$ are the strengths of 
the real and the imaginary parts, respectively, and $a_\text{R}$ and $a_\text{I}$ are their corresponding ranges.

It is worth recalling that the real part of the $\alpha$-interaction can be also derived via folding model calculations 
based on nucleon-nucleon interaction~\cite{lassaut}. When the antisymmetrization between the nucleons of the target and the 
nucleons from the $\alpha$ particle is taken into account, the exchange term makes the $\alpha$-interaction non-local. This 
non-local interaction can be replaced by a local interaction which depends on the energy. In Ref.~\cite{lassaut} it is shown 
that this energy-dependent interaction can be fitted by
\beq
v_{eff}(s,E) \approx U_0(1-\beta E) e^{-(s/a_\text{R})^2} .
\eeq
The parameters derived in~\cite{lassaut} for the Brink and Boeker force are $U_0=-47.3$~MeV and $\beta=0.003$~MeV$^{-1}$, 
while $a_\text{R}=1.97$~fm. This energy dependence is included in the present calculations.

The effective $\alpha$-nucleon interaction is also expected to depend on the nuclear matter density in which the nucleons 
interact. This density dependence is commonly represented in the following form ~\cite{satchler,kassem}:
\beq
f(\rho) = 1-\gamma \rho^{2/3},
\eeq
where $\gamma$ is a fitting parameter. In folding calculations, the factor $f(\rho)$ reduces the interaction strength in the interior of the target, where the density is highest, while leaving the surface and asymptotic regions largely unaffected. Such a modification is expected to improve the description of $\alpha$ scattering at large angles~\cite{satchler}.

Regarding the imaginary part of the $\alpha$-nucleon interaction, no fully reliable microscopic derivation is available to guide the construction of an effective interaction appropriate for ALAS studies. For example, a microscopic treatment of the imaginary interaction has been presented in Ref.~\cite{lassaut_i}; however, it relies on approximations that are not valid at the low incident energies relevant for ALAS. For this reason, in the present work we assume that the imaginary part of the 
$\alpha$-nucleon interaction exhibits the same functional dependence on energy and density as the real part.

Based on the above considerations, in this work we adopt the following parametrization for the $\alpha$-nucleon interaction:
\beq
v_{\alpha n} (s) = g_\text{R} f_\text{R}(E_\alpha,\rho) e^{-(s/a_\text{R})^2} + \mathrm{i} g_\text{I} f_\text{I}(E_\alpha,\rho) e^{-(s/a_\text{I})^2},
\eeq
where the density $\rho$ is evaluated at the position of the nucleon. The energy and density dependence 
is taken in the form 
\beq
f_k (E_\alpha,\rho) = U_0 (1-\beta E_\alpha) (1-\gamma_k \rho^{2/3}),
\eeq
where $k=\text{R},\text{I}$. Since the antisymmetrization effects are expected to influence the real and the imaginary 
parts in a similar manner, the same parameters $U_0$ and $\beta$ are used for both components.

\subsection{The nuclear density}

In the folding model [Eq.~\eqref{eq:folding_potential}], we employ target densities obtained from self-consistent relativistic and 
nonrelativistic mean-field calculations. In the following, we briefly outline the underlying framework of these models and discuss 
the resulting density distributions relevant for the present study.

\newpage

\noindent
{\it B1. Nuclear densities from relativistic mean-field calculations}
\vskip 0.2cm

These densities are computed within the relativistic Hartree-Bogoliubov (RHB) framework~\cite{ring}, which provides a unified 
description of particle-hole (ph) and pairing correlations. In the present calculations, the ph channel is described using the DD-PC1 
energy density functional~\cite{dd-pc1}, while pairing correlations are included in the neutron-neutron and proton-proton channels. 
The latter are treated using a separable Gaussian interaction adjusted to reproduce the pairing gap in nuclear matter obtained with 
the Gogny D1S force~\cite{duguet,tian}.

The RHB equations are solved in a basis of axially symmetric harmonic-oscillator wave functions. In this framework, the intrinsic ground states do not possess well-defined angular momentum and parity. To restore these symmetries, we apply the Generator Coordinate Method (GCM) and perform Hill-Wheeler-type configuration mixing calculations. Specifically, states with good
angular momentum and parity are constructed as:
\begin{equation}
\label{eq:HW_projection}
|J M \pi ; \alpha \rangle = \sum_i \sum_K f^{J K \pi}_ \alpha ( q_i) P^J_{MK} P^\pi |\phi (q_i) \rangle,
\end{equation}
where $P^J_{MK}$ and $P^\pi$ denote the angular-momentum and parity projection operators, respectively. 
This expression shows that the ground state $|J M \pi ; \alpha \rangle$ is a linear combination of 
projected RHB (proj-HFB) states, $P^J_{MK} P^\pi |\phi (q_i) \rangle$, each built from an intrinsic wave function
$|\phi (q_i) \rangle$ characterised by a quadrupole deformation $q_i$.
For convenience, we referred to this wave function as RHB+PGCM. Further details on the RHB+PGCM calculations can be found in Ref.~\cite{petar}.

\begin{figure}[htbp]
\centering
\includegraphics[width=\textwidth]{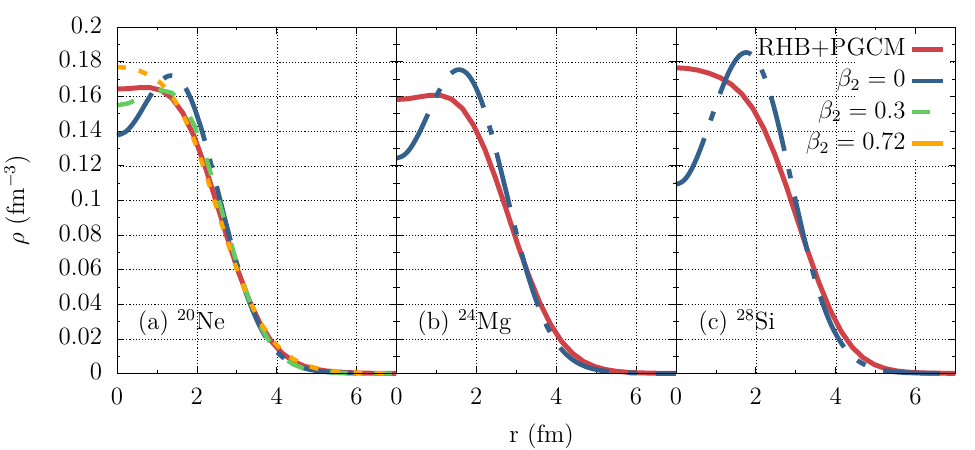}
\caption{\footnotesize Nuclear densities of $^{20}$Ne, $^{24}$Mg, and $^{28}$Si calculated within the RHB+PGCM framework and with spherically symmetric RHB solutions ($\beta_2 =0$). For $^{20}$Ne, densities calculated with proj-HFB states corresponding
to quadrupole deformations $\beta_2 = 0.3$ and $0.72$ are also shown.}
\label{fig:density_RMF}
\end{figure}

In Fig.~\ref{fig:density_RMF}, we display the densities obtained from RHB+PGCM calculations for the even-even $N = Z$ nuclei in the first half of the $sd$ shell. All 
considered nuclei exhibit deformed ground states, with quadrupole deformation parameters $\beta_2 = 0.27$, $0.43$, and $-0.38$, respectively. These values are 
obtained by averaging the contributions of the intrinsic $|\phi(q_i)\rangle$ states of different deformations that are mixed in the RHB+PGCM wave function. The 
corresponding experimental deformation parameters are $0.72$, $0.606$, and $0.412$, respectively. It should be noted that, for $^{28}$Si, experimental data cannot 
distinguish between prolate and oblate shapes.
 
For comparison, Fig.~\ref{fig:density_RMF} also includes the RHB densities calculated under the assumption of spherical symmetry. A pronounced difference between the 
RHB+PGCM and spherical RHB densities is observed, particularly in the nuclear interior, where the spherical RHB densities are significantly reduced. This 
behavior can be traced back to the weak pairing correlations in the spherical RHB calculations, which do not lead to a substantial occupation of the $2s_{1/2}$ 
single-particle orbital --- a state that contributes prominently to the central density.

In contrast, in the RHB+PGCM calculations, the occupation of the $2s_{1/2}$ orbital is enhanced due to the proximity of single-particle levels arising from the 
deformation-induced splitting of the $1d_{5/2}$ orbital. This results in an increased central density. To further illustrate the impact of deformation on the 
interior density, Fig.~\ref{fig:density_RMF} also shows, for $^{20}$Ne, the densities provided by the proj-HFB states
corresponding to the deformations $\beta_2 = 0.3$ and $0.72$. 

\vskip 0.2cm
\noindent
{\it B2. Nuclear densities from non-relativistic mean field calculations}
\vskip 0.2cm

The non-relativistic mean field is described with the quark-meson coupling (QMC) energy density functional (EDF)~\cite{qmc3}, which 
reproduces nuclear properties with an accuracy comparable to, and in some cases exceeding, that of standard Skyrme EDFs, while 
employing fewer adjustable parameters. In contrast to conventional mean-field approaches, where nucleons are treated as point-like 
particles, the QMC model describes nucleons as composite systems of three confined quarks interacting via scalar and vector meson 
fields. A notable feature of the QMC EDF is that its density dependence arises microscopically, while the spin-orbit and tensor terms 
are obtained without introducing additional parameters.

In the present QMC EDF calculations both the isovector ($T=1$) and isoscalar ($T=0$) pairing correlations
are included and treated within the quartet condensation model (QCM)~\cite{qcm_t0t1_nez}. 
In this framework, the ground state of even-even $N=Z$ nuclei is described by the wave function
\begin{equation}
   \label{eq:qcm}
   |\mathrm{QCM}\rangle=(Q^{\dag}_{T=1}+\Delta_0^{\dag 2})^{(N+Z)/4}|0\rangle,
\end{equation}
where $Q^{\dag}_{T=1}=2\Gamma_{1}^{\dag}\Gamma_{-1}^{\dag}-(\Gamma_{0}^{\dag})^{2} $ is the isovector quartet operator expressed in terms of the collective pair 
operators. The latter are defined by $\Gamma_{t}^{\dag}=\sum_{i}x_{i}P_{i,t}^{\dag}$, where $P_{i,t}^{\dag}$ creates an isovector pair in the time-reversed states 
$(i,\bar{i})$. The isoscalar pairing correlations are described by the collective isoscalar pair $\Delta_{0}^{\dag}=\sum_{i}y_{i}D_{i,0}^{\dag}$, where 
$D_{i,0}^{\dag}=(\nu_i^{\dag}\pi_{\bar{i}}^{\dag}-\pi_i^{\dag}\nu_{\bar{i}}^{\dag})/\sqrt{2}$. Finally, $n_q = (N+Z)/4$, where $N$ and $Z$ are the number of neutrons 
and protons above the core $|0\rangle$ which are considered in the pairing calculations. In the present QCM calculation we take $n_q =3$ (for details, see Ref.~\cite{qmc_qcm}).

The QCM wave function exhibits several distinctive features compared to the RHB+PGCM state discussed above [Eq.~\eqref{eq:HW_projection}]. Specifically: (i) it includes not 
only neutron-neutron and proton-proton pairing, but also both isovector and isoscalar proton-neutron pairing correlations; (ii) it exactly conserves particle number 
and isospin, although not angular momentum; and (iii) it incorporates four-body correlations induced by the pairing interaction.

The QCM calculations are done with the following isovector pairing interaction, derived within 
the QMC framework \cite{qmc3}
\begin{equation}
V_P^{(T=1)}(\mathbf{r},\mathbf{r}') = - s V_0 \left( 1- \eta \frac{\rho(\mathbf{r})}{1+d'G_\sigma \rho(\mathbf{r})} \right) 
\delta(\mathbf{r} - \mathbf{r}'),
\end{equation}
where $\rho$ is the baryon density, $V_0=G_\sigma-G_\omega - G_\rho/4 $ and $\eta =d'G^2_\sigma/V_0$. The interaction depends on the coupling constants for the 
scalar, vector and isovector mesons, $G_i={g_i^2}/{m_i^2}$, and on $d' = d + G_\sigma \lambda_3/3$, where $d$ is the nuclear polarizability and $\lambda_3$ is 
associated with the cubic term of the $\sigma$ field. For these parameters we employ the values provided in Ref.~\cite{qmc3}. Finally, $s$ is a scaling factor chosen to improve the description of odd-even mass differences for
$N > Z $ nuclei (for details, see Ref.~\cite{qmc_qcm})

For the isoscalar channel, the pairing interaction is assumed to be proportional to the isovector one, 
\begin{equation}\label{ISStrength}
    V_P^{(T=0)}=w V_P^{(T=1)}, 
\end{equation}
with $w=1.6$. This value lies in the range for which the Gamow-Teller strength for $^{56}$Ni, which depends only on the $T=0$ pairing, is reasonably reproduced \cite{gamow_teller}. 

The ground states of even-even $N = Z$ nuclei are determined by performing axially deformed QMC+QCM calculations, following the procedure outlined in Ref.~\cite{qmc_qcm}.
From the resulting axially symmetric densities, spherical densities are constructed by averaging over the polar angle,
\beq
   \label{eq:average_density}
   \rho(r) = \frac{1}{2} \int \limits_{0}^{\pi} \rho(r,\theta) \sin \theta \mathrm{d} \theta.
\eeq
This procedure provides an approximate restoration of spherical symmetry for the densities. A more rigorous approach would involve angular-momentum projection, as 
employed in the RHB+PGCM calculations discussed in the previous subsection. However, such a method is not currently available within the QMC+QCM framework.

\begin{figure}[htbp]
\centering
\includegraphics[width=\textwidth]{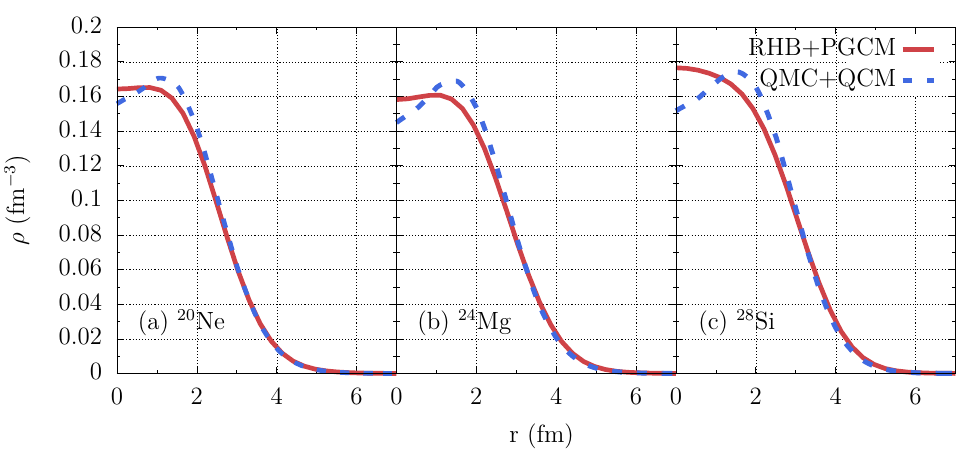}
\caption{\footnotesize Nuclear densities of $^{20}$Ne, $^{24}$Mg, and $^{28}$Si calculated  within
the QMC+QCM approach, compared with the RHB+PGCM densities.}
\label{fig:density}
\end{figure}

Fig.~\ref{fig:density} shows the densities obtained from the QMC+QCM calculations for $^{20}$Ne, $^{24}$Mg, and $^{28}$Si, whose 
intrinsic quadrupole deformations are $0.474$, $0.478$, and $-0.308$, respectively. The corresponding RHB+PGCM densities are also 
included in the figure for comparison. It can be observed that the QMC+QCM densities are lower in the central region and larger at 
radial distances between approximately 1 and 2--3~fm. This behavior is mainly due to the reduced occupation of the $2s_{1/2}$ orbital 
in the QMC+QCM approach compared to RHB+PGCM, which in turn enhances the population of deformed states originating from the 
$1d_{5/2}$ orbital. These states are more localized at finite radii inside the nucleus, leading to a depletion of the central density 
and an enhancement in the region $r \approx 1-3$~fm.

As a general conclusion, the densities in the nuclear interior exhibit a strong sensitivity to the choice of mean field, pairing interaction, and its treatment. 
However, this sensitivity is significantly reduced in the surface and tail regions, which are expected to play the dominant role in the low-energy 
$\alpha$-scattering processes analyzed in this work.

\subsection{Evaluation of the $\alpha$-particle elastic differential cross section}

The differential cross section has been evaluated with the nuclear potential provided by Eq.~\eqref{eq:folding_potential}. 
The latter has been obtained by numerical integration in spherical coordinates. The calculation has been optimized by 
assuming the $z$-axis aligned with the distance $r_\alpha$ between the $\alpha$ particle and the target nucleus. With this choice, the dependence on the azimuthal angle $\varphi$ is eliminated. The nuclear potential can then be written as
\begin{equation}
    V_{\alpha n}(r_\alpha) = 2 \pi \int \limits_0^{R_\mathrm{nuc}} \int \limits_0^{\pi}
                               \rho(r_n) \hspace{0.05cm} v_{\alpha n}(s) r_n^2
                               \sin \theta \hspace{0.05cm} \mathrm{d}r_n \hspace{0.05cm} \mathrm{d}\theta,
\end{equation}
where $s$ denotes the distance between the $\alpha$ particle and the nucleon,
\begin{equation}
\label{eq:s}
    s = \sqrt{r_\alpha^2 + r_n^2 + 2 \hspace{0.04cm} r_\alpha \hspace{0.055cm} r_n \cos \theta}.
\end{equation}

With the total nuclear and the electrostatic potential we solve the radial Schrödinger equation:
\begin{equation}
\label{eq:radial_eq}
  -\frac{\hbar^2}{2\mu} \frac{d^2 P_{E \ell}(r)}{dr^2}
  + \left[ \frac{\hbar^2 \ell (\ell + 1)}{2 \mu r^2} + V(r) \right] P_{E \ell}(r)
  = E P_{E \ell }(r),
\end{equation}
where $\ell$ is the orbital angular momentum, and $\mu$ is the reduced mass
\begin{equation}
\label{eq:mu}
\mu = \frac{m_{\alpha} m_\mathrm{t}}{m_{\alpha} + m_\mathrm{t}},
\end{equation}
with $m_{\alpha}$ and $m_\mathrm{t}$ denoting the masses of the $\alpha$ particle and the target nucleus, 
respectively. The Schrödinger equation has been solved numerically using the RADIAL Fortran subroutine 
package~\cite{radial} and the phase shift $\delta_{\ell}$ was derived for each orbital angular momentum $\ell$.

Finally, the differential cross section for the elastic scattering of an $\alpha$ particle along a direction
$\hat{\boldsymbol{\Omega}} = (\theta, \varphi)$ with respect to the incident direction, is expressed as:
\begin{equation}
\label{eq:dxsalpha}
    \cfrac{\mathrm{d} \sigma}{\mathrm{d} \hat{\boldsymbol{\Omega}}}
     =\big|f(\hat{\boldsymbol{\Omega}})\big|^2 ,
\end{equation}
where the scattering amplitude has the standard expression
\begin{equation}
\label{eq:ampl}
\begin{aligned}
    f(\hat{\boldsymbol{\Omega}}) = \frac{1}{2 \mathrm{i} k} \sum_{\ell=0}^{\infty}(2 \ell+1)
       P_{\ell}(\cos \theta)\left(\mathrm{e}^{2 \mathrm{i} \delta_{\ell}}-1\right) .
\end{aligned}
\end{equation}
In this expression, $k$ represents the center-of-mass wavevector, and $P_{\ell}(\cos \theta)$ are the Legendre polynomials. 

\subsection{Calculation scheme and fitting procedure}

The differential cross section (DCS) depends on the inputs used in the folding calculations, namely the nuclear densities and the $\alpha$-nucleon interaction. The parameters of the $\alpha$-nucleon interaction are determined through the fitting procedure described below. In this procedure, we employ nuclear densities obtained from the RHB+PGCM calculations described above. 
This choice is motivated by the fact that, within the RHB+PGCM framework, spherical symmetry is properly restored through angular-momentum projection.

To reduce the number of free parameters, the energy-dependent factor of the $\alpha$-nucleon interaction is taken from the theoretical estimates of Ref.~\cite{lassaut}. In addition, the range of the real part is fixed at $a_\text{R} = 2$~fm, which is close to the value $a_\text{R} = 1.97$~fm considered in Ref.~\cite{lassaut}.

\begin{table}[h!]
\caption{\footnotesize Minimum and maximum values of each parameter, together with the corresponding number of evaluation points, used in the initial coarse fitting step.}
\label{tab:sweeping_param}
\centering
\begin{tabular}{| c | c | c | c | c | c |}
\hline
Parameter & $g_\mathrm{R}$ & $g_\mathrm{I}$ & $a_\mathrm{I}$ (fm) & $\gamma_\mathrm{R}$ (fm$^2$) & $\gamma_\mathrm{I}$ (fm$^2$) \\
\hline
Min. & $0.1$ & $0.01$ & $1.9$ & $1.7$ & $1.7$ \\
\hline
Max. & $3.5$ & $1.0$ & $2.3$ & $2.1$ & $2.1$ \\
\hline
$N_p$ & $8$ & $7$ & $9$ & $5$ & $5$ \\
\hline
\end{tabular}
\end{table}

The rest of the parameters have been determined through a fitting procedure as follows. For each pair of target nucleus and 
$\alpha$-particle energy considered in this work, the fit parameters have been varied initially on a coarse grid, as 
displayed in Table~\ref{tab:sweeping_param}. Then, the interaction potential, Eq.~\eqref{eq:total_pot}, has been evaluated 
for each combination of fit parameters. Next, the DCS has been calculated for each evaluated 
interaction potential, as discussed in the foregoing section. 

\begin{table}[h!]
\centering
\resizebox{\textwidth}{!}{%
\begin{tabular}{|c|c|c|c|c||c|c|c|c|c|}
\hline
Nucleus & $E_{\alpha}$ (MeV) & $g_\mathrm{R} \pm \sigma_{g_\mathrm{R}}$ & $g_\mathrm{I} \pm \sigma_{g_\mathrm{I}}$ & Cor($g_\mathrm{R}$,$g_\mathrm{I}$) &
Nucleus & $E_{\alpha}$ (MeV) & $g_\mathrm{R} \pm \sigma_{g_\mathrm{R}}$ & $g_\mathrm{I} \pm \sigma_{g_\mathrm{I}}$ & Cor($g_\mathrm{R}$,$g_\mathrm{I}$) \\
\hline
\multirow{4}{*}{$^{20}$Ne}
 & $33.00$  & $2.7 \pm 0.26$ & $0.30 \pm 0.21$ & $-0.34$ &
\multirow{4}{*}{$^{40}$Ar} 
 & $22.10$  & $2.0 \pm 0.037$ & $0.20 \pm 0.026$ & $-0.213$ \\
 & $50.90$  & $2.0 \pm 0.71$ & $0.45 \pm 0.26$ & $0.54$  &
 & $24.10$  & $2.1 \pm 0.097$ & $0.20 \pm 0.063$ & $-0.216$ \\
 & $54.10$  & $2.1 \pm 0.68$ & $0.50 \pm 0.30$ & $0.53$  &
 & $29.20$  & $1.9 \pm 0.192$ & $0.25 \pm 0.116$ & $-0.275$ \\
 & $104.00$ & $2.7 \pm 0.93$ & $0.80 \pm 0.30$ & $0.84$  &
 & $104.00$ & $2.1 \pm 0.108$ & $0.70 \pm 0.046$ & $0.702$  \\
\hline
\multirow{4}{*}{$^{24}$Mg} 
 & $22.10$  & $2.0 \pm 0.068$ & $0.15 \pm 0.004$ & $0.34$  &
\multirow{4}{*}{$^{40}$Ca} 
 & $22.00$  & $2.1 \pm 0.648$ & $0.10 \pm 0.0029$ & $0.90$ \\
 & $39.00$  & $2.7 \pm 0.036$ & $0.30 \pm 0.019$ & $0.014$ &
 & $24.10$  & $2.1 \pm 0.094$ & $0.10 \pm 0.0098$ & $0.91$ \\
 & $50.00$  & $2.7 \pm 0.086$ & $0.45 \pm 0.016$ & $0.068$ &
 & $29.00$  & $2.2 \pm 0.052$ & $0.15 \pm 0.0053$ & $0.03$ \\
 & $119.70$ & $1.5 \pm 0.060$ & $0.50 \pm 0.029$ & $0.90$  &
 & $104.00$ & $1.5 \pm 0.815$ & $0.45 \pm 0.0228$ & $0.79$ \\
\hline
\multirow{4}{*}{$^{28}$Si} 
 & $40.00$  & $1.8 \pm 0.064$ & $0.25 \pm 0.023$ & $-0.20$ &
\multirow{4}{*}{$^{44}$Ca} 
 & $22.00$  & $1.8 \pm 0.033$ & $0.15 \pm 0.0129$ & $0.496$ \\
 & $45.00$  & $1.3 \pm 0.054$ & $0.30 \pm 0.017$ & $0.34$  &
 & $24.10$  & $1.9 \pm 0.022$ & $0.15 \pm 0.0109$ & $0.806$ \\
 & $50.00$  & $2.3 \pm 0.062$ & $0.35 \pm 0.015$ & $0.24$  &
 & $29.00$  & $2.2 \pm 0.023$ & $0.20 \pm 0.0199$ & $0.556$ \\
 & $104.00$ & $2.2 \pm 0.079$ & $0.60 \pm 0.023$ & $0.77$  &
 & $104.00$ & $1.4 \pm 0.050$ & $0.50 \pm 0.0244$ & $0.747$ \\
\hline
\multirow{4}{*}{$^{36}$Ar} 
 & $17.50$  & $2.2 \pm 0.020$ & $0.09 \pm 0.002$ & $0.061$ &
\multirow{4}{*}{$^{48}$Ca}
 & $22.00$  & $2.1 \pm 0.051$ & $0.15 \pm 0.0264$ & $0.473$ \\
 & $22.10$  & $2.2 \pm 0.032$ & $0.15 \pm 0.010$ & $0.002$ &
 & $24.10$  & $2.1 \pm 0.041$ & $0.15 \pm 0.0073$ & $-0.146$ \\
 & $24.10$  & $2.2 \pm 0.041$ & $0.15 \pm 0.014$ & $0.455$ &
 & $29.00$  & $1.7 \pm 0.016$ & $0.15 \pm 0.0099$ & $0.788$ \\
 & $29.20$  & $1.9 \pm 0.016$ & $0.15 \pm 0.007$ & $0.448$ &
 & $104.00$ & $2.1 \pm 0.060$ & $0.60 \pm 0.0193$ & $0.790$ \\
\hline
\end{tabular}
}
\caption{\footnotesize Fit parameters, their standard deviations $\sigma$, and their correlation coefficients, for the target nuclei and $\alpha$-particle energies considered in this work.}
\label{tab:fit_param}
\end{table}

To maximize the agreement between the distorted-wave and the experimental DCS, the following measure has been adopted:
\begin{equation}
\label{eq:chi2_alpha}
    \chi^2 = \displaystyle \sum_{i=1}^{N}
             \left(
             \log{\cfrac{\mathrm{d}\sigma (\theta_{i})}{\mathrm{d}\sigma_\mathrm{R}} \Bigg \vert_\mathrm{DW}}
             -
             \log{\cfrac{\mathrm{d}\sigma (\theta_{i})}{\mathrm{d}\sigma_\mathrm{R}} \Bigg \vert_\mathrm{exp}}
             \right)^2
             ,
\end{equation}
where $\cfrac{\mathrm{d}\sigma (\theta_{i})}{\mathrm{d}\sigma_\mathrm{R}} \Bigg \vert_\mathrm{DW}$ is the distorted-wave DCS 
at scattering angle $\theta_{i}$, $\cfrac{\mathrm{d}\sigma (\theta_{i})}{\mathrm{d}\sigma_\mathrm{R}} \Bigg 
\vert_\mathrm{exp}$ is the experimental DCS~\cite{exfor,zerkin} at $\theta_{i}$, both expressed as a ratio to the Rutherford 
DCS, and $N$ is the number of experimental CM scattering angles. The combination of fit parameters which minimizes 
Eq.~\eqref{eq:chi2_alpha} is searched. Once this is found, the fitting procedure is restarted, employing a refined grid 
around the best found fit parameters.

Following the procedure described above, the parameters $a_\mathrm{I}=2.25$~fm, $\gamma_\mathrm{R}=1.8$~fm$^2$, and 
$\gamma_\mathrm{I}=1.7$~fm$^2$ were found to provide a good description for all the nuclei and all $\alpha$-particle energies 
considered in this work. Thus, the only remaining energy- and nucleus-dependent parameters are the renormalization factors 
$g_\mathrm{R}$ and $g_\mathrm{I}$ of the strengths of the real and of the imaginary parts, respectively, of the 
$\alpha$-nucleon interaction. These are reported in Table~\ref{tab:fit_param}, along with their correlation coefficient and 
their standard deviations, which were calculated  using the covariance matrix, as explained in 
the Appendix.

\section{Results and Discussions}

\subsection{Differential cross sections with RHB+PGCM densities}
\label{subsec:accurate}

\begin{figure}[h!]
\centering
\includegraphics[width=0.925\textwidth]{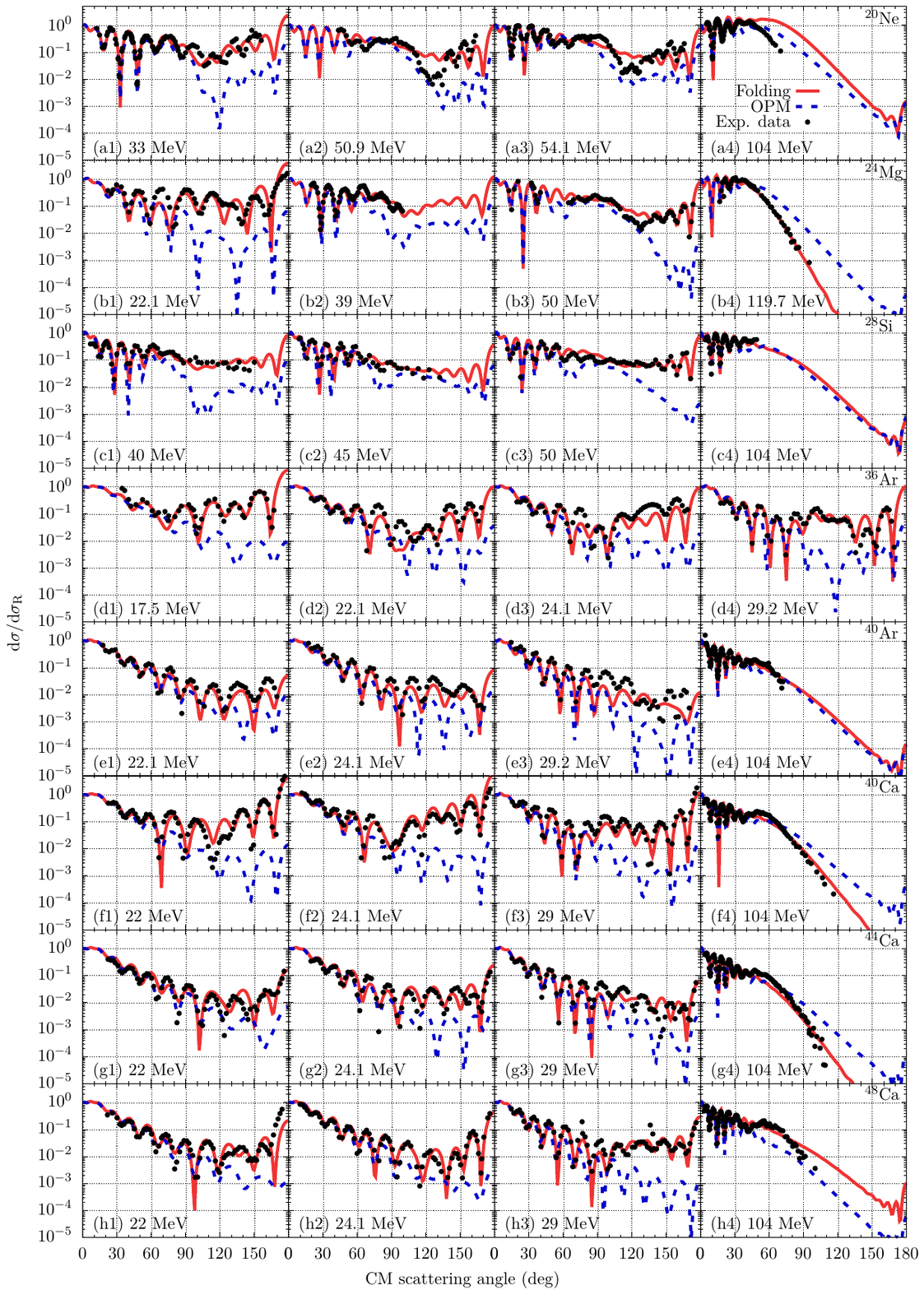}
\caption{\footnotesize Comparison of DCSs obtained with the single-folding model employed in this work (solid red curves) and those calculated using a standard global OPM~\cite{su} (dashed blue curves), for selected nuclei and $\alpha$-particle energies. Experimental angular distributions~\cite{exfor,zerkin} are shown as black dots.}
\label{fig:su}
\end{figure}

\begin{figure}
\centering
\includegraphics[width=0.6\columnwidth]{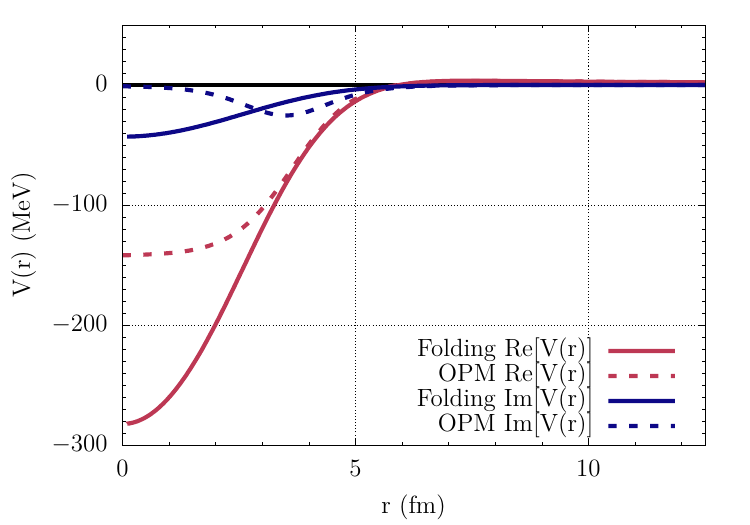}
\caption{\footnotesize Interaction potentials (real and imaginary) for the scattering 
of $33$~MeV $\alpha$ particles on $^{20}$Ne.}
\label{fig:pot_alpha}
\end{figure}

The main results for the DCSs are presented in Fig.~\ref{fig:su}, which compares 
partial-wave DCSs obtained with the single-folding potential (solid red curves) to those calculated using a standard global optical model (OPM)~\cite{su} (dashed blue curves) for selected target nuclei and $\alpha$-particle energies. Experimental angular distributions~\cite{exfor,zerkin} are shown as black dots. The single-folding results are computed using the $\alpha$-nucleon interaction specified above and nuclear densities provided by the  RHB+PGCM calculations.

As shown in Fig.~\ref{fig:su}, at low energies --- where ALAS is typically observed --- the global OPM significantly 
underestimates the experimental DCS at intermediate and large scattering angles. In contrast, the single-folding calculations 
reproduce the pronounced backscattering features observed in the data. Moreover, as seen from Fig.~\ref{fig:su}, the same set of 
parameters provides a good description of the DCS at higher scattering energies for all calculated nuclei, with the exception of 
$^{20}$Ne.

The large differences between the DCSs obtained with the global OPM and the single-folding approach are also
reflected in the corresponding interaction potentials. As an illustrative example, Fig.~\ref{fig:pot_alpha} 
shows the interaction potential for the scattering of $33$~MeV $\alpha$ particles on $^{20}$Ne. While the real
parts of the two potentials exhibit similar radial dependence, their depths differ significantly.
In contrast, the imaginary parts display pronounced differences both in magnitude and shape. In particular, 
substantial deviations are observed in the central region as well as in the nuclear surface.

\vskip 0.2cm
\noindent
{\it A1. Differential cross sections for neutron-rich nuclei }
\vskip 0.2cm

\begin{figure}[h]
\centering
\includegraphics[width=\textwidth]{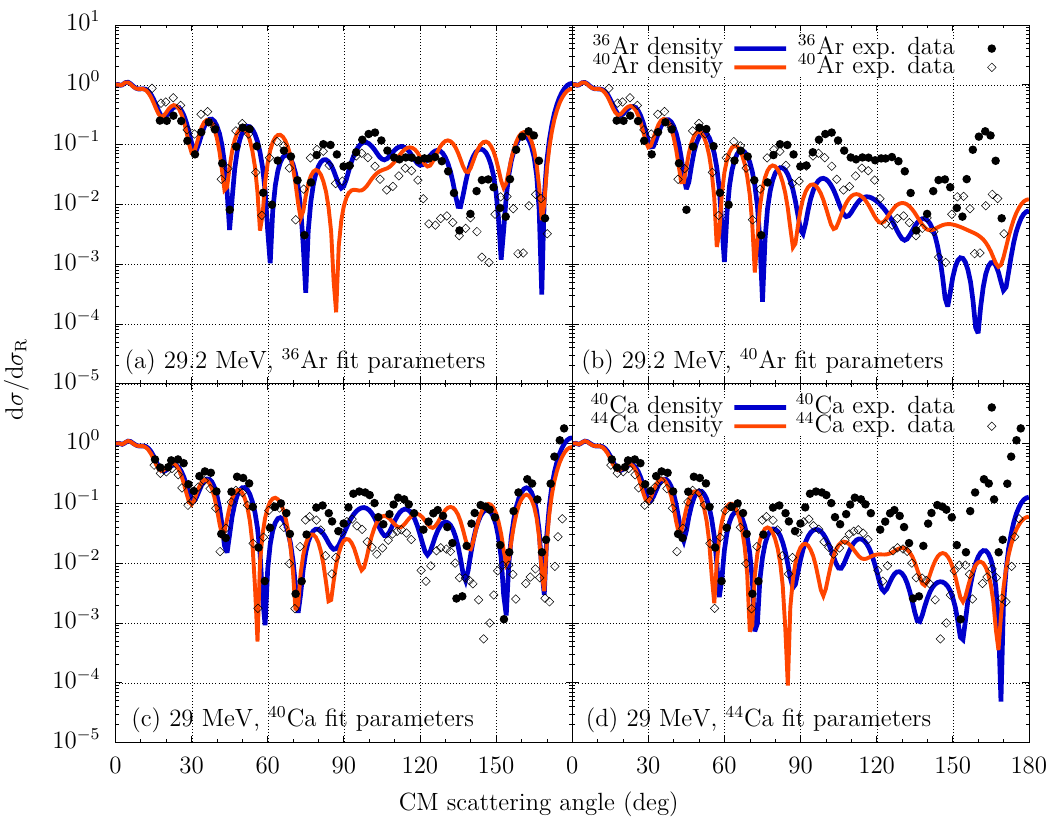}
\caption{\footnotesize Cross sections for the elastic scattering of $\alpha$ 
particles on Ar and Ca isotopes. See text for details.}
\label{fig:n_excess}
\end{figure}

Experimental angular distributions show that the enhancement at large scattering angles is more pronounced in $N=Z$ nuclei than in neutron-rich systems. This 
behaviour is evident in Fig.~\ref{fig:n_excess}, where the experimental DCSs~\cite{exfor,zerkin} for $^{36,40}$Ar and $^{40,44}$Ca are presented 
(black dots). Folding-model calculations based on microscopic densities satisfactorily reproduce the dependence of the large-angle DCS on neutron excess.

In the folding calculations, different values of the parameters $g_\text{R}$ and $g_\text{I}$ are used for $N=Z$ and $N>Z$ nuclei 
(Table~\ref{tab:fit_param}). To isolate the effect of density changes associated with the additional neutrons, we evaluate the DCS by 
interchanging the parameter sets between $N=Z$ and $N>Z$ isotopes. The results for $^{36,40}$Ar and $^{40,44}$Ca at $\alpha$-particle 
energies of 29.2~MeV and 29~MeV, respectively, are shown in Fig.~\ref{fig:n_excess}. At these energies, $g_\text{R}$ is the same for both 
classes of nuclei (Table~\ref{tab:fit_param}), allowing a more direct assessment of density effects.

A substantial modification of the DCS is observed when the densities of $N=Z$ and $N>Z$ nuclei are interchanged. However, changing 
only the densities while keeping the imaginary strength $g_\text{I}$ fixed does not provide a satisfactory description of the data. 
For instance, as shown in Fig.~\ref{fig:n_excess}(b), the DCS for $^{36}$Ar cannot be reproduced from that of $^{40}$Ar by simply 
replacing the $^{40}$Ar density with that of $^{36}$Ar while retaining the $g_\text{I}$ value used for $^{40}$Ar. A reasonable 
agreement is achieved only by reducing $g_\text{I}$ from 0.25 to 0.15 (Table~\ref{tab:fit_param}).

These results indicate that ALAS in $N=Z$ nuclei cannot be accounted for solely by ground-state density effects. Instead, they point 
to a significant role of reaction dynamics associated with the excitation spectrum, which is effectively incorporated through the 
imaginary part of the optical potential.

In fact, the low-energy excitation spectra of $^{36}$Ar and $^{40}$Ar, which contribute to the inelastic channels included in the 
imaginary term, differ markedly. In $^{36}$Ar, there are only four excited states below 4.5~MeV: the first at 1.970~MeV, followed by 
three states above 4~MeV. By contrast, $^{40}$Ar exhibits 24 states in the same energy region, with the lowest at 1.460~MeV and three 
additional states below 3~MeV. This difference in the excitation spectra is consistent with the folding-model results, which require 
a weaker imaginary potential for $^{36}$Ar than for $^{40}$Ar.

%\begin{figure}
%\centering
%\includegraphics[width=0.75\textwidth]{density_Ar_Ca.pdf}
%\caption{\footnotesize Comparison between nuclear densities of Ar (left panel) and Ca (right panel) isotopes.}
%\label{fig:density_Ar_Ca}
%\end{figure}

\vskip 0.2cm
{\it A2. Effect of nuclear deformations on differential cross sections}
\vskip 0.2cm

Most $N=Z$ nuclei exhibit ground-state deformation. As the density distribution can change significantly with deformation (see, \textit{e.g.}, Fig.~\ref{fig:density_RMF}), a 
corresponding impact on the DCS is expected. To illustrate this effect, we consider the nucleus $^{20}$Ne, which is strongly deformed in its ground state, with 
$\beta_2 = 0.72$.

\begin{figure}
\centering
\includegraphics[width=1.0\textwidth]{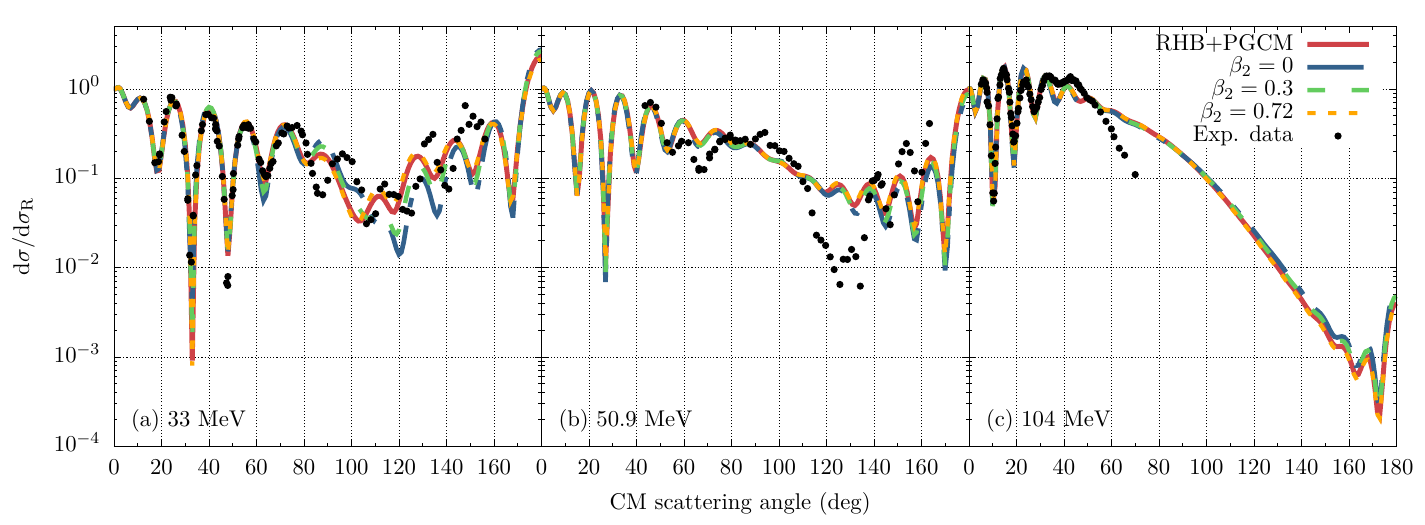}
\caption{\footnotesize Effect of deformation in $^{20}$Ne on the differential cross section of $\alpha$-particle at different energies.}
\label{fig:20Ne_deformations}
\end{figure}

Figure~\ref{fig:20Ne_deformations} shows the DCS calculated for a range of deformations, from the spherical case up to $\beta_2 = 
0.72$. The RHB+PGCM solution corresponds to a mixing of states with different deformation, with an average of $\beta_2 = 0.27$. For 
low-energy $\alpha$ scattering at 33~MeV, the DCS exhibits a strong dependence on deformation in the angular range 
$100^\circ$-$140^\circ$. In particular, at $\theta = 120^\circ$, the DCS increases by nearly an order of magnitude as the deformation 
changes from $\beta_2 = 0$ to $\beta_2 = 0.72$.

In contrast, at higher scattering energies the sensitivity of the DCS to deformation becomes much weaker. These results indicate that an accurate treatment of 
nuclear deformation is essential for describing low-energy scattering, where ALAS typically emerges.

\subsection{Differential cross sections with  QMC+QCM densities}
\label{subsec:density_effect}

\begin{figure}
\centering
\includegraphics[width=\textwidth]{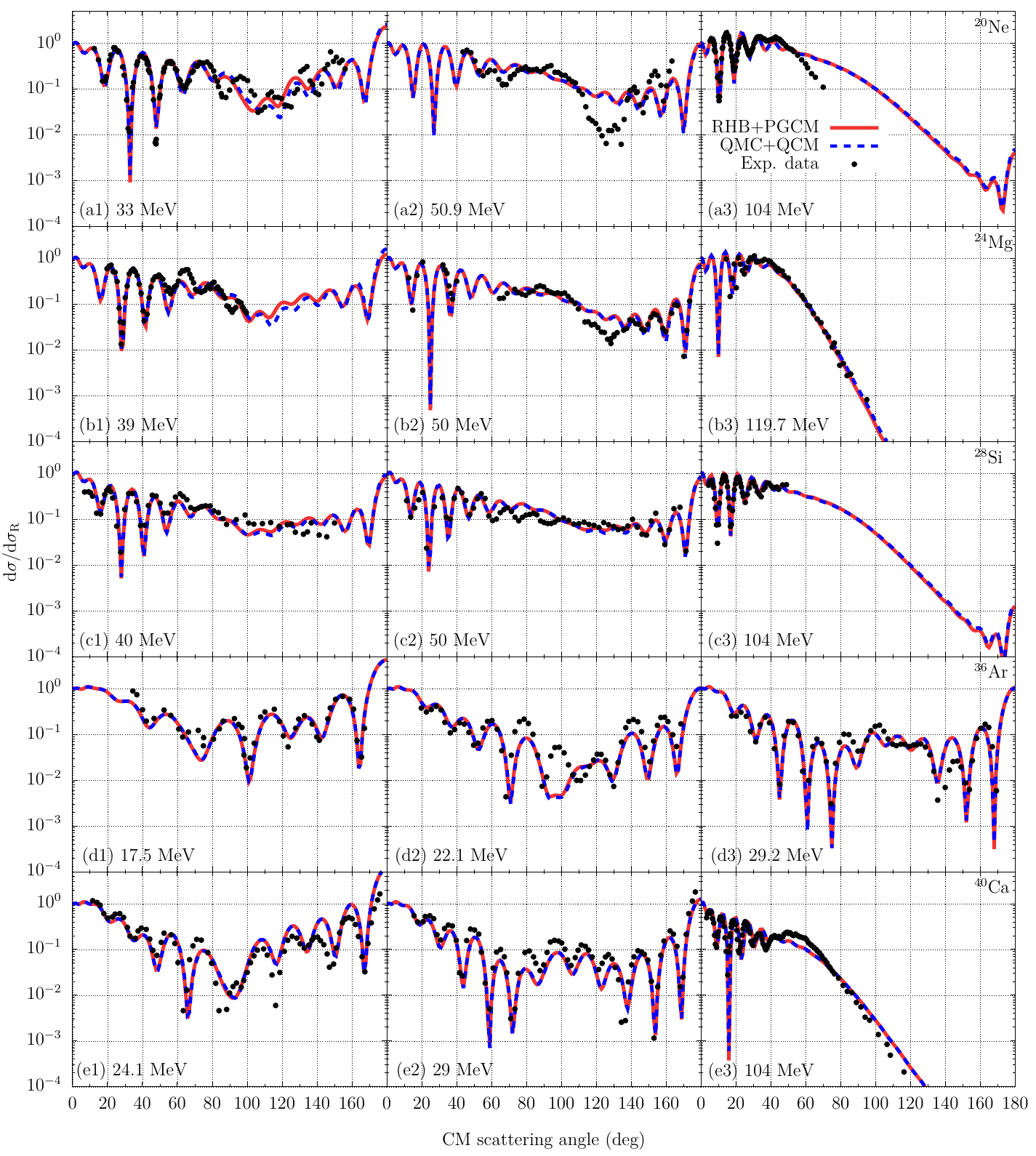}
\caption{\footnotesize Differential cross sections calculated with the QMC+QCM densities, compared with 
results obtained with  RHB+PGCM densities. Experimental angular distributions~\cite{exfor,zerkin} are shown as black dots.}
\label{fig:qmc}
\end{figure}

To compute the DCSs using the QMC+QCM densities, a new set of $\alpha$-nucleon interaction parameters should, in principle, be determined following the same fitting 
procedure adopted for the RHB+PGCM densities. However, in order to assess the sensitivity of the DCSs to the choice of nuclear densities employed in the single-folding 
model, we perform calculations using the parameter set listed in Table~\ref{tab:fit_param}. The most relevant results are presented in Fig.~\ref{fig:qmc}.

Overall, the DCSs obtained with the two sets of densities are similar. Although the QMC+QCM densities differ substantially from the RHB+PGCM ones in the nuclear interior 
(see, \textit{e.g.}, Fig.~\ref{fig:density}), the corresponding DCSs exhibit only minor variations. This indicates that, in low-energy elastic collisions, the $\alpha$ 
particle predominantly probes the nuclear surface, where the two density distributions differ only slightly.

The most noticeable deviations occur for $^{20}$Ne at $E_{\alpha} = 33$~MeV and for $^{24}$Mg at $E_{\alpha} = 39$~MeV, around scattering angles of $120^\circ$.

These differences are most likely related to the distinct deformations predicted by the RHB+PGCM and QMC+QCM approaches. However, 
a direct comparison between these deformations is not straightforward. In the QMC+QCM framework, the deformation characterizes an axially symmetric intrinsic state, 
whereas in the RHB+PGCM approach it is associated with a spherically symmetric  state constructed as a superposition of intrinsic configurations with different 
deformations.

The observation that, in some cases, the DCS in the ALAS region is significantly influenced by nuclear deformation (see also 
Fig.~\ref{fig:20Ne_deformations}) is noteworthy. Since deformation originates from dynamical effects associated with ground-state 
correlations --- such as the interplay between mean-field, pairing and four-body correlations --- the DCS in low-energy 
$\alpha$ scattering on $N=Z$ nuclei in the ALAS region may provide a sensitive probe of these effects. A dedicated investigation of 
this possibility, however, is beyond the scope of the present work.

\section{Summary and Conclusions}

In this study, we have shown that a single-folding model based on nuclear densities derived from self-consistent mean-field 
calculations is able to describe reasonably well the ALAS in even-even $sd$-shell nuclei. In the present framework, we employ an 
energy- and density-dependent nucleon-$\alpha$ interaction of Gaussian form for both the real and imaginary components of the 
scattering potential. The range of the interaction and its energy dependence are constrained by previous theoretical studies. Within 
this approach, a satisfactory description of the experimental data is achieved using only two adjustable parameters, namely the 
strengths of the real and imaginary parts of the interaction, $g_\text{R}$ and $g_\text{I}$.

Our analysis indicates that, in order to reproduce ALAS, smaller values of $g_\text{I}$ are required for $N=Z$ nuclei than for $N>Z$ 
systems. In particular, we find that simply replacing the density of an $N>Z$ nucleus with that of an $N=Z$ nucleus, without reducing 
$g_\text{I}$, is insufficient to reproduce ALAS in $N=Z$ nuclei. Since $g_\text{I}$ is associated with the effects of inelastic 
channels, this result suggests that ground-state properties --- encoded in the nuclear density --- are not, by themselves, sufficient to 
account for ALAS.

One of the longstanding open questions in ALAS studies is why this phenomenon occurs predominantly in even-even $N=Z$ nuclei and at low scattering energies. An early interpretation attributed ALAS to the scattering of $\alpha$ particles from preformed $\alpha$ clusters assumed to exist at the surface of even-even $N=Z$ nuclei~\cite{schmeing,michel_alpha,mahmoud}. In such approaches, the interaction is effectively treated as scattering between two $\alpha$ particles, implying that the clusters behave as localized $\alpha$-like objects. However, this assumption is not supported by microscopic calculations, including the QCM approach employed here.
In $N=Z$ nuclei, $\alpha$-like structures are more appropriately described as four-body correlations in spin and isospin, which are not necessarily spatially localized.

These four-body correlations have a significant impact on binding energies~\cite{qcm_binding,qmc_qcm} and can also modify the ground-state density. However, as shown in Sec.~III~B, such modifications do not lead to appreciable changes in the folding-model results. This suggests that ALAS is more closely related to the excitation properties of even-even $N=Z$ nuclei, which are influenced by $\alpha$-like correlations induced by proton-neutron pairing~\cite{qcm_ex1,qcm_ex2}.

More specifically, low-energy excitations in even-even $N=Z$ nuclei require the breaking of four-body correlations (\textit{i.e.}, $\alpha$-like quartets), followed by the breaking of neutron-neutron, proton-proton, and/or proton-neutron pairs. By contrast, in even-even $N>Z$ nuclei, low-energy excitations predominantly involve pair breaking only. As a consequence, low-energy excitations occur more readily in $N>Z$ nuclei than in $N=Z$ nuclei.

For example, in $^{20}$Ne there are only eight excited states below 7~MeV, with the lowest at 1.633~MeV, whereas in $^{22}$Ne there are 18 states in the same energy region, with the lowest at 1.274~MeV. A similar trend is observed in the case of $^{36}$Ar and $^{40}$Ar (see Sec.~III~A1).

Consequently, the incident low-energy $\alpha$-particle flux is expected to experience weaker absorption in even-even $N=Z$ nuclei. 
Within the single-folding model, this reduced absorption --- closely related to ALAS~\cite{michel1,michel2} --- is 
incorporated phenomenologically through a smaller imaginary component of the $\alpha$-nucleon interaction in $N=Z$ systems. A fully 
microscopic description of this mechanism would require evaluating the absorption in terms of the excitation spectra of $N=Z$ nuclei; 
however, such an approach remains highly challenging and lies beyond the scope of the present work.

\vskip 0.2cm
\noindent
{\bf Acknowledgments}
\vskip 0.2cm
\noindent

N. S. acknowledges the support from CERN during his visit related to this study. He also acknowledges the funding from the Romanian Ministry of Education 
and Research through the project PN-23-21-01-01/2023. The work of P.M. was funded by the European Union’s Horizon Europe research and innovation program under the Marie Skłodowska-Curie Actions Grant Agreement No. 101149053.

\section*{APPENDIX}

For a model that depends on $n$ parameters, $\boldsymbol{p}=\{p_1,p_2,...,p_n\}$, the best fit parameters, 
$\boldsymbol{p}_0$, can be found by minimizing
\begin{equation}
   \chi^2(\boldsymbol{p}) = \displaystyle \sum \limits_{i=1}^{N} \left( y_i^{\mathrm{exp}} - 
                            y_i^{\mathrm{eval}}(\boldsymbol{p}) \right)^2, 
\end{equation}
where $N$ is the number of experimental data points, $y_i^{\mathrm{exp}}$ are the experimental quantities at each 
experimental data point $i$, and $y_i^{\mathrm{eval}}(\boldsymbol{p})$ are the corresponding fitted quantities at each 
experimental data point $i$, depending on the fit parameters $\boldsymbol{p}$.

After finding the best-fit parameters $\boldsymbol{p}_0$ by minimizing $\chi^2(\boldsymbol{p})$, the covariance matrix of
the parameters can be evaluated as~\cite{covariance}:
\begin{equation}
   \boldsymbol{C} = \frac{\chi^2(\boldsymbol{p}_0)}{N-n} \left( \boldsymbol{J}^{T} \boldsymbol{J} \right)^{-1},
\end{equation}
where $N-n$ is the number of degrees of freedom and $\boldsymbol{J}$ is the Jacobian matrix, with dimensions $N \times 
n$. This matrix consists of the first-order derivatives of the residuals, $r_i(\boldsymbol{p}) = y_i^{\mathrm{exp}} - 
y_i^{\mathrm{eval}}(\boldsymbol{p})$, with respect to the parameters:
\begin{equation}
   J_{ij} = \frac{\partial r_i(\boldsymbol{p})}{\partial p_j}.
\end{equation}
When an explicit analytical form of the derivatives is not available, finite differences can be used to approximate them:
\begin{equation}
   J_{ij} = \frac{\partial r_i(\boldsymbol{p})}{\partial p_j}
          \approx \frac{r_i(\boldsymbol{p_0} + \Delta \boldsymbol{p_j}) - r_i(\boldsymbol{p_0})}{\Delta p_j}.
\end{equation}

Once the covariance matrix $\boldsymbol{C}$ is known, the correlation coefficient $\rho_{ij}$ between the parameters $p_i$ 
and $p_j$ is
\begin{equation}
   \rho_{ij} = \frac{C_{ij}}{\sqrt{C_{ii}C_{jj}}},
\end{equation}
where $C_{ij}$ is the covariance between $p_i$ and $p_j$, and $C_{ii}$ and $C_{jj}$ are their variances.

The correlation coefficients can take values between $-1$ and $1$:
\begin{itemize}
   \item $\rho_{ij} = -1$ indicates a perfect negative correlation, \textit{i.e.}, when one parameter increases, the other 
   decreases.
   \item $\rho_{ij} = 0$ shows that there is no correlation between the parameters.
   \item $\rho_{ij} = 1$ indicates a perfect positive correlation, \textit{i.e.}, when one parameter increases, the other 
   increases as well.
\end{itemize}

\newpage
%% Bibliography

\end{document}